\documentclass[a4paper]{article}
\usepackage[medium,compact]{titlesec}
\usepackage{INTERSPEECH2020}
\usepackage{makecell}
\usepackage{subfigure}
\usepackage{cite}
\title{Unsupervised Cross-Lingual Speech Emotion Recognition Using Domain Adversarial Neural Network}
\name{Xiong Cai$^1$, Zhiyong Wu$^{1,2}$, Kuo Zhong$^1$,  Bin Su$^1$, Dongyang Dai$^1$, Helen Meng$^{1,2}$}
\address{
    $^1$Tsinghua-CUHK Joint Research Center for Media Sciences, Technologies and Systems,\\Shenzhen International Graduate School, Tsinghua University, Shenzhen, China \\ $^2$Department of Systems Engineering and Engineering Management, \\The Chinese University of Hong Kong, Shatin, N.T., Hong Kong SAR, China
}
\email{\{cai-x18,zhongk17,sub18,ddy17\}@mails.tsinghua.edu.cn, \{zywu,hmmeng\}@se.cuhk.edu.hk}

\begin{document}

\maketitle
\begin{abstract}
    By using deep learning approaches, Speech Emotion Recognition (SER) on a single domain has achieved many excellent results. However, cross-domain SER is  still a challenging task due to the distribution shift between source and target domains. In this work, we propose a Domain Adversarial Neural Network (DANN) based approach to mitigate this distribution shift problem for cross-lingual SER\@. Specifically, we add a language classifier and gradient reversal layer after the feature extractor to force the learned representation both language-independent and emotion-meaningful. Our method is unsupervised, i.\,e., labels on target language are not required, which makes it easier to apply our method to other languages. Experimental results show the proposed method provides an average absolute improvement of 3.91\% over the baseline system for arousal and valence classification task. 
    Furthermore, we find that batch normalization is beneficial to the performance gain of DANN\@. Therefore we also explore the effect of different ways of data combination for batch normalization.
\end{abstract}

\section{Introduction}

With the extensive application of Artificial Intelligence (AI) products in our daily lives, it has become increasingly imperative to design a smarter Human-Computer Speech Interaction (HCSI) system. Speech Emotion Recognition (SER), which aims to infer the emotional state of a speaker from his or her speech \cite{r01}, has been regarded as a crucial component for a more intelligent HCSI system. Existing SER models \cite{r02,r03,r04} have achieved satisfactory level results when the training and test data are from the same corpus. However, it is still intractable to build a more robust cross-lingual SER system because of the domain shift between corpora of different languages \cite{r05}.

Numerous approaches have been proposed to reduce the domain shift problem for cross-corpus or cross-lingual SER\@. \cite{r06} proposes a fine-grained adversarial domain adaptation scheme, which reduces the distribution shift of the same emotion class in different corpora. \cite{r07} shows that fine-tuning can effectively improve the recognition results. These methods are promising, but additional labeled data are required, which might not be available since their collection is expensive.  

A more practical solution is unsupervised domain adaptation which only demands unlabeled data from related domains. A number of previous studies have explored statistical-based methods to reduce  mismatch between domains \cite{r08, r09, r10, r11, r12}. Specifically, \cite{r08,r09} deploy different level of feature normalization strategies to minimize the speaker-and-corpus-related effects; \cite{r10, r11, r12} apply the Maximum Mean Discrepancy (MMD) or Kernel Canonical Correlation Analysis (KCCA) approaches to increase the similarity or correlation of different domains. All these methods reduce the domain shift directly on the original input feature space or its linear transformation space, so the capacity of shift-reduction might be limited. Some other studies \cite{r13, r14, r15} use variants of autoencoder to learn a concise and common feature representation by incorporating the prior knowledge from unlabeled data into learning. Since the optimization of the autoencoder and emotion classifier is not performed simultaneously, it is not clear whether compressed representations preserve all the emotion information of speech.

Recently, Adversarial Learning (AL), such as Generative Adversarial Network (GAN) \cite{r16} and Domain Adversarial Neural Network (DANN) \cite{r17}, has become an increasingly popular approach for domain adaptation. \cite{r19} proposes a GAN-based model for cross-lingual SER  and  demonstrates significant improvements even for the non-mainstream Urdu language. \cite{r20, r21} use a DANN-based framework to learn a speaker-independent representation and greatly improve the single-corpus results. \cite{r22} explores the advantage of DANN for cross-corpus SER on three English corpora. Besides, the DANN techniques have also been widely applied in other speech applications such as automatic speech recognition \cite{r23} and speaker recognition \cite{r26} to deal with the domain mismatch problem.

Inspired by the success of DANN in domain adaptation tasks, this paper proposes a DANN-based approach to reduce the distribution shift for cross-lingual SER\@. Specifically, based on the primary emotion classification task, a language classifier with Gradient Reversal Layer (GRL) is added to the model as an auxiliary task to help learn language-independent representations. Unlike the studies mentioned above, our approach reduces the distribution shift in a compressed feature space instead of the original input space, and all the modules are trained jointly rather than separately, which makes the model learn emotion-discriminative and language-independent representations more efficiently. Our contribution is two-fold: First, we introduce the DANN framework for cross-lingual SER and achieve significant performance improvements. Second, our study presents that batch normalization (BN) \cite{r27} can contribute to improve DANN and explores four different ways of combining data for BN\@. 

The rest of this paper is organized as follows. The proposed method is described in Section 2. The databases and classification scheme are detailed in Section 3. Experimental setup and results analysis are presented in Section 4. Section 5 finalizes the study with conclusions and future directions.

\section{Methodology}
Firstly, we formulate our cross-lingual SER as the following domain adaptation task. We have a source language corpus with emotion labels as source domain, $D_s=\{(\bm{x}_i^s,\bm{y}_i^s)\}_{i=1}^{n}$, and a target language corpus without emotion labels as target domain, $D_t=\{\bm{x}_i^t\}_{i=1}^m$, where $\bm{x}_i^s$ and $\bm{x}_i^t\in \mathbb{R}^{k\times d}$, $\bm{y}_i^s\in\{0,1\}^c$; $k$, $d$ and $c$ are the number of frames in an utterance, the dimension of each feature frame and the number of emotion categories; $n$ and $m$ are the number of samples in $D_s$ and $D_t$. Our goal is to learn a reliable emotion classifier from the labeled $D_s$ and the unlabeled $D_t$, which can be generalized well in $D_t$.

\subsection{Model structure}
\begin{figure}[t]
  \centering
    \includegraphics[width=\linewidth]{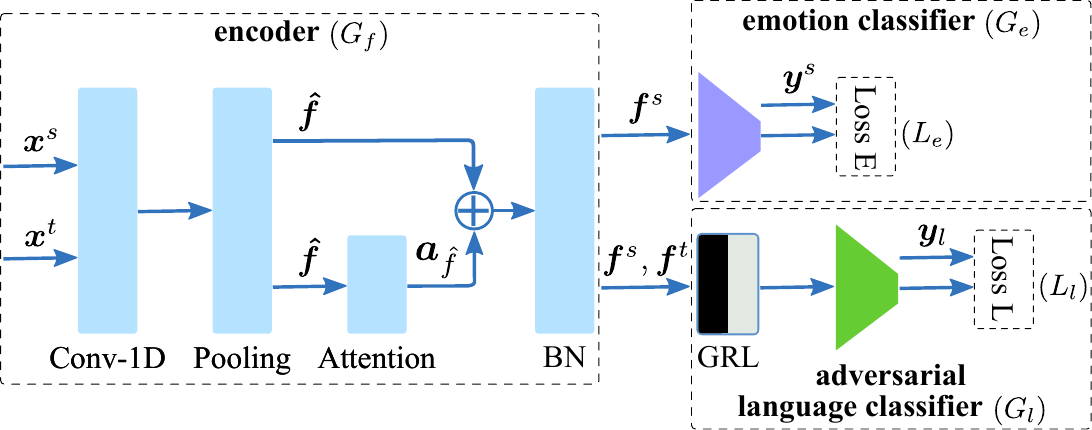}
  \caption{The proposed model structure. BN and GRL are Batch Normalization, and Gradient Reversal Layer respectively.} 
  \label{fig:1}
\end{figure}

As shown in Figure~\ref{fig:1}, the proposed model consists of three modules: encoder ($G_f$), emotion classifier ($G_e$) and language classifier ($G_l$). The encoder structure is mainly adopted from \cite{r07},  except that we add a batch normalization  (BN) layer for the stability of training a DANN model. A 1D convolution layer with ReLU activation takes Mel spectral features as input to capture emotion-related patterns. Then, a max pooling layer with a large stride follows to select the most salient features. Next, an attentive vector $\bm{a}_{\hat{f}}$ is extracted from the outputs of max pooling by the following attention formulas:
\begin{align}
    s_i & = \frac{\exp(\bm{v}^T\bm{\hat{f}}_i)}{\sum\limits_j \exp(\bm{v}^T\bm{\hat{f}}_j)} \\
    \bm{a}_{\hat{f}} & = \sum \limits _{i} s_i \bm{\hat{f}} _i
\end{align}
where $\bm{\hat{f}}_i$ is the $i$-th feature vector of the output $\bm{\hat{f}}$ of max pooling layer and $\bm{v}$ is a trainable vector as a global attention query. The motivation behind using this attention mechanism is that emotion-related information is distributed differently over the utterance. This global attention query $\bm{v}$ can be used to learn to capture these important emotion patterns.  Finally, the attention vector $\bm{a}_{\hat{f}}$ is appended to the end of the output $\bm{\hat{f}}$ of max pooling along the time dimension, and then all these feature vectors are flattened into a fixed-length vector as the input of the following BN layer. As the final representation, the output of the BN layer $\bm{f}$ is fed into emotion classifier (only source domain data) and language classifier (both source and target domain data). As for classifiers, a single dense layer with softmax activation and two output units are used for both classifiers. Besides, a Gradient Reversal Layer (GRL) is inserted between BN layer and language classifier to achieve the goal of adversarial training.

\subsection{Adversarial training}
DANN \cite{r17} is a fairly elegant neural network framework for unsupervised domain adaptation, where unlabeled target domain data can be efficiently utilized to reduce the variations between the source and target domains. Specifically, there are two tasks: a primary target task (e.\,g., emotion classification) and an auxiliary domain classification task (e.\,g., language classification). Both tasks share the feature extractor and a GRL is introduced between feature extractor and domain classifier. GRL is a layer without trainable parameters and works as a “pseudo-function” $R(\bm{x})$ defined as the following formulas:
\begin{align}
R(\bm{x})&=\bm{x} \\
\frac{\mathrm{d}R}{\mathrm{d}\bm{x}}&=-\beta\bm{I}
\end{align}
where $\bm{I}$ is an identity matrix and $\beta$ is a hyper parameter controlling the scale of reversal gradient signal. By using GRL, the trainable parameters before and after GRL are updated in the opposite direction, namely, \textit{adversarial training}. As for the classifiers of the two tasks, parameters are updated to minimize their respective errors. As for the feature extractor, parameters are updated to minimize the error of primary task while maximizing the error of domain classification task, where the latter is implemented by GRL\@. Therefore, the learned feature could be meaningful for the primary task and indistinguishable for the domain classifier. In terms of our cross-lingual SER, the primary task is emotion classification and the auxiliary task is language classification. Our goal is to learn a representation that retains discriminative information for emotion and reduces variations for languages. Therefore, the learned feature extractor and emotion classifier can be directly applied to target language data.

We use the cross-entropy loss as the training objective for both emotion and language classifiers:
\begin{align}
L_e(\bm{\theta}_f,\bm{\theta}_e) & =-\frac{1}{n}\sum_{(\bm{x},\bm{y})\in D_s} \bm{y}^T\log G_e(G_f(\bm{x};\bm{\theta}_f); \bm{\theta}_e) \\
L_l(\bm{\theta}_f,\bm{\theta}_l) & =-\frac{1}{n+m}\sum_{\bm{x}\in D} \bm{y}_l^T\log G_l(G_f(\bm{x};\bm{\theta}_f); \bm{\theta}_l)
\end{align}
where\;$D=D_s\cup D_t$; $L_e$ and $L_l $ are the losses for emotion and language classifiers respectively; $\bm{\theta}_f$, $\bm{\theta}_e$  and $\bm{\theta}_l$ represent the trainable parameters for $G_f$, $G_e$ and $G_l $ respectively;  $\bm{y}$ is  the one-hot encoding for emotion labels from source domain and $\bm{y}_l $ is the one-hot encoding for language labels distinguishing source and target languages. 

Then, we define the total loss as the weighted sum of the above two losses and directly minimize it for training:
\begin{align}
L(\bm{\theta}_f,\bm{\theta}_e, \bm{\theta}_l)=\alpha\cdot L_e(\bm{\theta}_f,\bm{\theta}_e)+(1-\alpha)\cdot L_l(\bm{\theta}_f,\bm{\theta}_l)
\end{align}
where $\alpha$ plays a trade-off for the two losses. Due to the existence of GRL, minimizing the total loss $L$ will actually lead to the following way of parameter update:
\begin{align}
\bm{\theta}_e &\leftarrow\bm{\theta}_e-\lambda\cdot\alpha\frac{\partial L_e}{\partial \bm{\theta}_e}\\
\bm{\theta}_l &\leftarrow\bm{\theta}_l-\lambda\cdot(1-\alpha)\frac{\partial L_l}{\partial \bm{\theta}_l}\\
\bm{\theta}_f &\leftarrow\bm{\theta}_f-\lambda\cdot\bigg(\alpha\frac{\partial L_e}{\partial\bm{\theta}_f}-(1-\alpha)\cdot\beta\frac{\partial L_l}{\partial \bm{\theta}_f}\bigg)
\end{align}
Concretely, $\bm{\theta}_e$ and $\bm{\theta}_l$  are updated for minimizing  $Le$ and $L_l$ respectively, and $\bm{\theta}_f$ is updated for minimizing $L_e$ while  maximizing  $L_l$ simultaneously. $\lambda$ and $\beta$ are the learning rate and the gradient reversal scale of  GRL\@. After training, a feature representation rich in emotional information and indistinguishable from languages will be obtained from the encoder output.

\section{Databases}
\subsection{IEMOCAP}
IEMOCAP \cite{r28} is an audiovisual database of English dyadic conversations performed by ten professional actors. There are two types of conversations: the scripted ones and the improvised ones (given a certain scenario and topic). This corpus contains a total of 10,039 utterances, where audio, video, text and motion-capture recordings are available. The categorical emotion label and 5-point scales on the dimensions valence, arousal, and dominance (1 - low/negative, 5 - high/positive) are annotated by at least 2 raters. In our study, only audio modal data and dimension label of valence and arousal are used.
\subsection{RECOLA}
RECOLA \cite{r29} is a multimodal database of French dyadic conversations. Participants express emotions spontaneously during a collaborative video conference. Four different modal data of audio, video, electrocardiogram (ECG) and electrodermal activity (EDA) are recorded continuously and synchronously. Continuous valence and arousal labels in the range[-1, 1] are measured by 6 annotators at frame level. Since our goal is to predict emotion on utterance level, the mean value across all frames of an utterance and all annotators are calculated as the final label. Freely available 1,308 audio utterances from 23 speakers are used in our study. 
\subsection{Classification scheme and input features}
In this work, we focus on a binary classification task of  valence (negative/positive) and arousal (low/high). In order to obtain binary training labels, we use the same annotation mapping scheme as in \cite{r07}. For IEMOCAP, the two ranges [1, 2.5] and (2.5, 5] are categorized as low/negative and high/positive respectively. Similarly, the corresponding two ranges are [-1, 0] and (0, 1] for RECOLA\@. In terms of input features, 26 logMel filter-banks are extracted frame-wise from a single utterance with frame size of 25ms and frame shift of 10ms. The logMel feature has a fixed length of 750 frames. The shorter one is padded with the minimum for each dimension in an utterance and the longer one is truncated to 750 frames in the middle. 


\section{Experiments}
\subsection{Experimental setup}
We use the following configurations for model training. 200 filters with kernel size 10 and stride 3 are used for the 1D convolution layer. The size and stride of max pooling are both  set to 30. Adam\cite{r30} optimizer and exponential decay learning rate with initial rate 1e-3, decay rate 0.93 for every epoch, and final rate 5e-5 are used to optimize parameters. For the regularization, dropout with rate 0.7 as suggested in \cite{r31} is used  for the output of encoder; $l_1$ and $l_2$  regularization with the weight 5e-3 are used for training RECOLA and IEMOCAP respectively. We train the models for 50 epochs with a batch size of 32, and 30\% of data from test set is used as the development set for early stopping. The logMel features are normalized with zero mean and unit variance for each database. All experiments are run five times with different random seeds, and the unweighted average recall (UAR) is chosen as our evaluation criterion.
\subsection{Experimental results}
\subsubsection{Performance of the proposed model}
In this section, we compare three trained models: our proposed model (\emph{\textbf{our}}), the baseline model (\emph{\textbf{base}}), and the mono-lingual model (\emph{\textbf{mono}}). The \emph{\textbf{base}} and \emph{\textbf{mono}} model use the structure which consists of the same encoder and emotion classifier only as in Figure~\ref{fig:1}. The \emph{\textbf{mono}} model is trained and tested on the same database, where 70\% samples are used for training, 25\% for testing and 5\% for early stopping. It provides us with an idea about the best achievable results within each database. We use \emph{\textbf{Rec}} and \emph{\textbf{Iem}} to represent the RECOLA and IEMOCAP database, and \emph{\textbf{Rec2Iem}} means training on \emph{\textbf{Rec}} and testing on \emph{\textbf{Iem}} and vice versa.

Table~\ref{tab:1} reports UAR (\%) results with standard deviations in parentheses for the three models. Comparing the results of \emph{\textbf{base}} and \emph{\textbf{our}}, our proposed model outperforms the baseline model in all experiments and achieves an average improvement of 3.91\%. This result presents that the proposed approach can effectively reduce variations between different languages while retaining the information related to emotions. Therefore, the emotion classifier can benefit from the learned language-independent representation to improve results in target language. To illustrate this, we use Principal Component Analysis (PCA) to project the learned feature representation, i.\,e., the output of encoder, into 2D space.
\begin{table}[th]
  \vspace{-0.2cm} 
  \caption{UAR (\%)  for baseline and  proposed method.}
  \label{tab:1}
  \centering
  \renewcommand\tabcolsep{1.3pt}
  \scalebox{0.845}{
  \begin{tabular}{cccccc}
    \toprule
    & \multicolumn{2}{c}{\textbf{Rec2Iem}} & \multicolumn{2}{c}{\textbf{Iem2Rec}} & \\
    \cmidrule(lr){2-3} \cmidrule(lr){4-5}
    \textbf{model} & \textbf{arousal} &\textbf{valence} & \textbf{arousal} & \textbf{valence} & \textbf{average} \\
    \midrule
    \textbf{base} & $62.49(2.96)$ & $54.15(0.51)$ & $60.73(0.45)$ & $58.11(0.51)$ & $58.87$ \\
    \textbf{our}  & $\bm{71.99}(\bm{0.33})$ & $54.54(0.77)$ & $\bm{63.18}(\bm{0.32})$ &  $\bm{61.43}(\bm{1.38})$ & $\bm{62.78}$ \\
    \textbf{mono}\footnotemark[1] & $75.55(0.78)$ & $63.20(2.23)$ & $66.28(1.71)$ & $62.90(1.12)$ & $66.98$ \\
    
    \bottomrule
  \end{tabular}}
  \vspace{-0.2cm} 
\end{table}
\footnotetext[1]{\textbf{mono} is trained and tested on \textbf{Iem} for \textbf{Rec2Iem} setup, and \textbf{Rec} for \textbf{Iem2Rec} setup} 

\begin{figure}[t]
 \centering
   \subfigure[Feature representation learned by the baseline model]{
   \label{fig:3a}
   \begin{minipage}[b]{0.46\textwidth}   
   \includegraphics[width=1\textwidth]{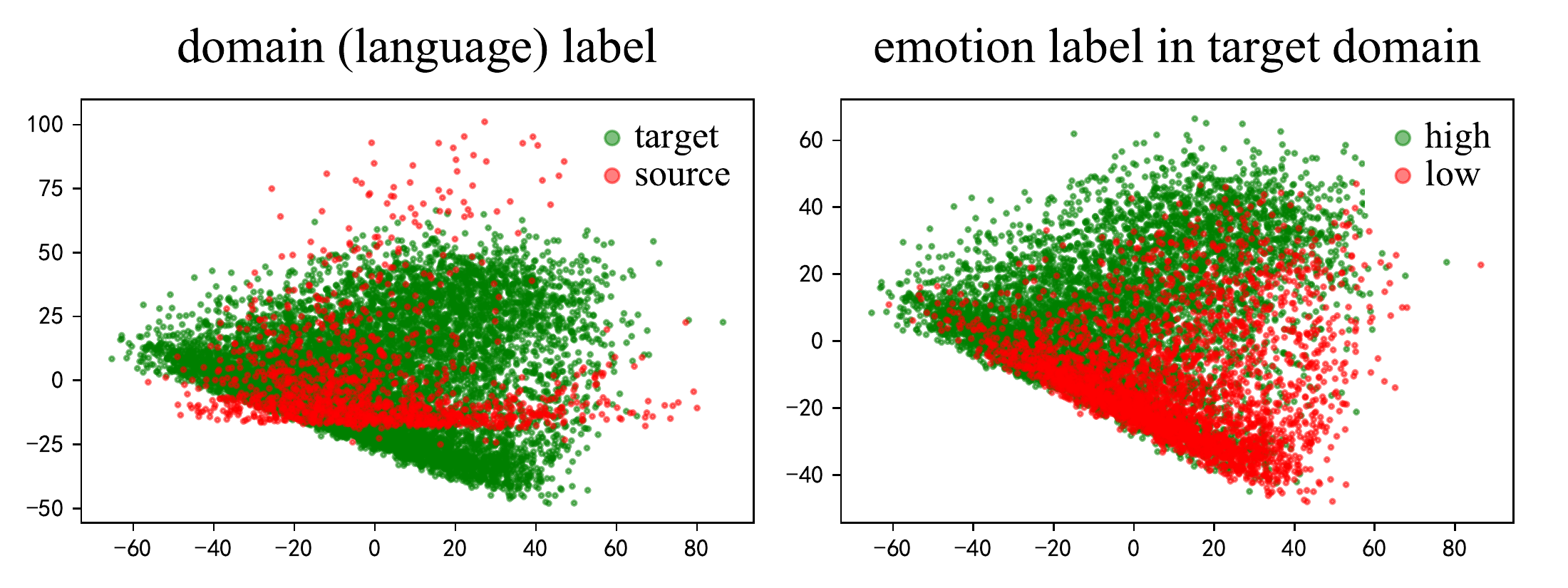}
   \end{minipage}}
   \subfigure[Feature representation learned by the proposed model]{
   \label{fig:3b}
   \begin{minipage}[b]{0.46\textwidth}
   \includegraphics[width=1\textwidth]{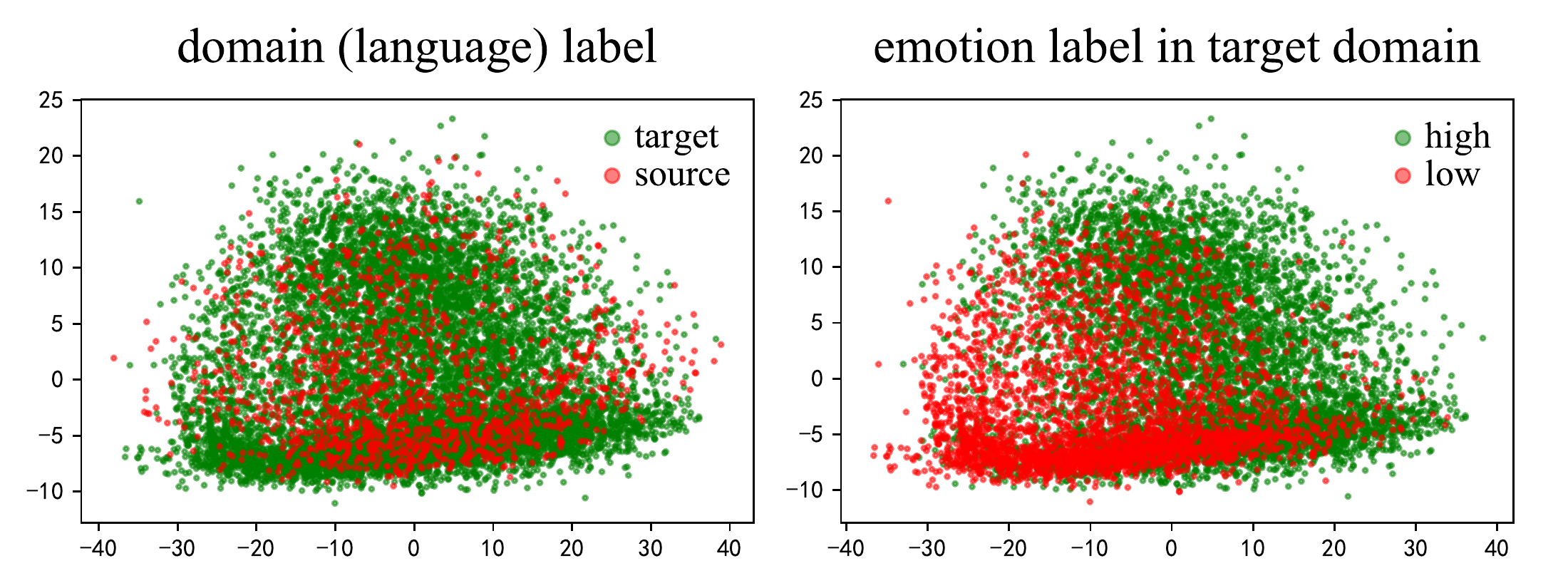}
   \end{minipage}}
 \caption{PCA plot of the learned feature representation with language labels (left) and emotion labels (right) for baseline and our proposed model from the “Rec2Iem arousal” training.}
 \label{fig:3}  
 \end{figure}

\begin{figure}[t]
  \centering
  \includegraphics[scale=0.35]{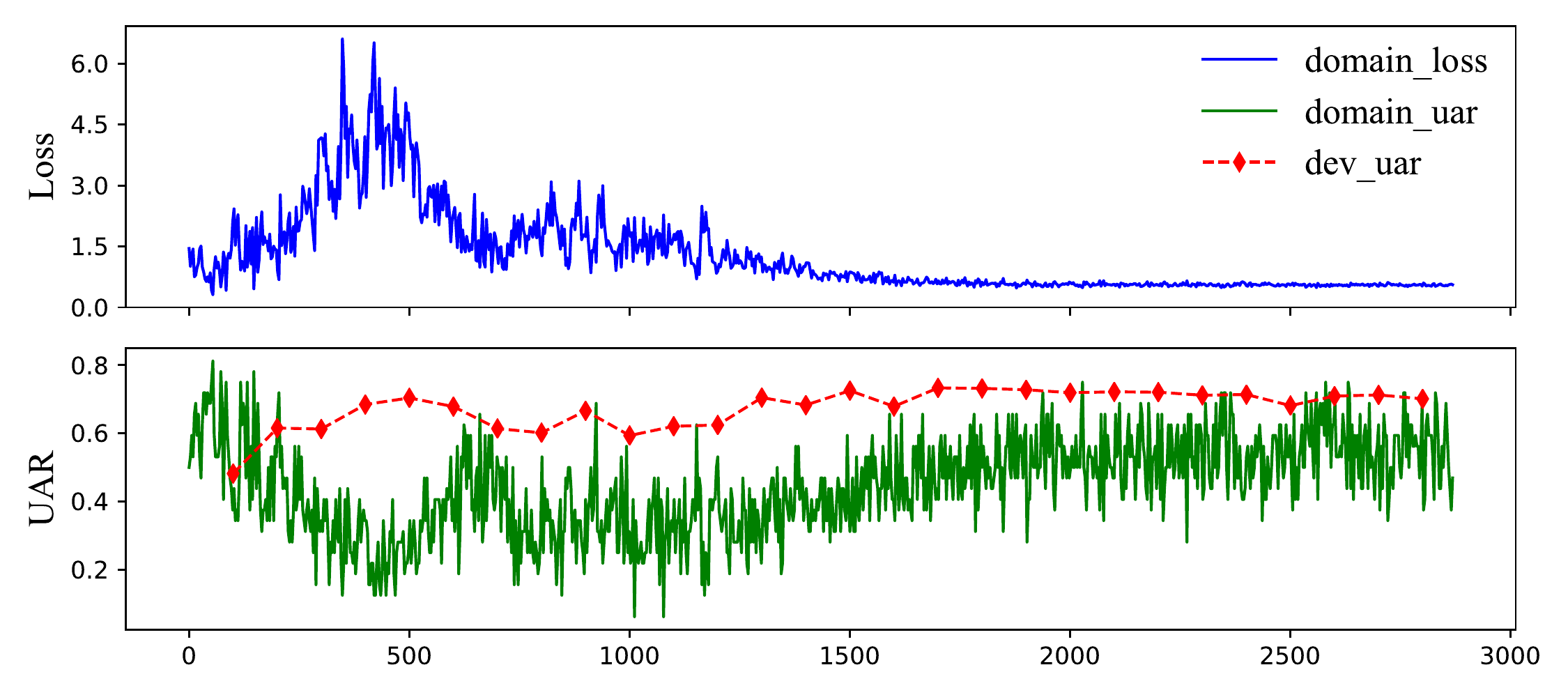}
  \caption{Domain(language) loss, domain UAR, and emotion development set UAR from the “Rec2Iem arousal" training} 
  \label{fig:4}
\end{figure}

As shown in Figure~\ref{fig:3}, regarding language labels, feature representations learned by our adversarial training model (Figure~\ref{fig:3b} left) are more evenly mixed and therefore more indistinguishable than the ones learned by baseline model (Figure~\ref{fig:3a} left); while, as for emotion labels, feature representations learned by  our proposed model (Figure~\ref{fig:3b} right) are more separable than the ones learned by baseline model (Figure~\ref{fig:3a} right). Besides, in Figure~\ref{fig:4}, we plot the training curves of the domain classification loss (blue line) and UAR (green line), and the emotion classification UAR (red line) in development set. It can be seen that, at the early stage of training, the domain loss increases and decreases alternately, and the domain UAR changes oppositely than it, which suggests that the adversarial training itself works well. Moreover, the emotion development set UAR has a similar trend with domain loss, which means that the emotion classifier gets better results when the domain classifier has a higher loss, i.\,e., the more language-indiscriminative the features are, the better performance of emotion classification will be. These visual results further indicate the effectiveness of the proposed method for cross-lingual SER\@.  

 Comparing the results of \emph{\textbf{base}} and \emph{\textbf{mono}} in Table~\ref{tab:1}, the performance of naive cross-lingual SER (baseline) is 8.11\% lower on average than the mono-lingual SER\@. This result consistent with \cite{r05, r32} indicates that the distribution shift between different languages will seriously harness the predictive ability of SER\@. In addition, another clear conclusion can be obtained that the prediction of arousal is easier than valence regardless of cross-lingual or mono-lingual tasks. Similar results can be found in \cite{r05, r33, r34, r35}. This is mainly because acoustic features such as energy, pitch and speed are related to arousal \cite{r01}, but for valence, there is no consensus on how acoustic features correlate with it, and it is more speaker-dependent \cite{r31}.

\subsubsection{Impact of  batch normalization on performance}
In this section, we first study the effect of BN on widening the performance gap between our DANN-based model and base- line model. As shown in Table~\ref{tab:2}, our proposed model is only 2.61\% higher than baseline on average when BN is not used, which is much lower than the gap of 3.91\% as in Table~\ref{tab:1} with BN used. A possible reason for this result is that the adversarial training of our model is not as stable as baseline, while BN can help to make the distribution of the representation layer more stable \cite{r27}. Therefore, better results can be achieved after using the BN layer in our adversarial training model.
\begin{table}[th]
  \vspace{-0.1cm} 
  \caption{UAR (\%)  for no batch normalization.}
  \label{tab:2}
  \centering
  \renewcommand\tabcolsep{1.3pt}  
  \scalebox{0.82}{
  \begin{tabular}{cccccc}
    \toprule
    & \multicolumn{2}{c}{\textbf{Rec2Iem}} & \multicolumn{2}{c}{\textbf{Iem2Rec}} & \\
    \cmidrule(lr){2-3} \cmidrule(lr){4-5}
    \textbf{model} & \textbf{arousal} &\textbf{valence} & \textbf{arousal} & \textbf{valence} & \textbf{average} \\
    \midrule
    \textbf{base}  & $58.04(1.90)$ & $52.34(0.79)$ & $58.69(0.95)$ & $53.79(0.87)$ & $55.71$ \\    
    \textbf{our}   & $\bm{63.62}(\bm{1.74})$ & $\bm{52.94}(\bm{0.36})$ & $\bm{59.76}(\bm{0.77})$ &  $\bm{57.81}(\bm{2.07})$ & $\bm{58.32}$ \\
    \bottomrule
  \end{tabular}}
  
\end{table}

Based on the conclusion reached above, we further explore four different ways of combining data for BN\@. For the training of DANN model, both source and target data need to be fed into the model. They can be first combined into one mini-batch and then fed into the model, or each of them occupy a min-batch and fed into the model alternatively. For the first data feeding method, three ways of combining data for BN are performed as follows: perform BN on the whole batch, namely \emph{\textbf{BN1}}, which is used for above experiments, where the first half batch (source half) is fed to the emotion classifier ($G_e$) and the whole batch is fed to the language classifier ($G_l$); perform BN on the source half batch and whole batch respectively, namely \emph{\textbf{BN2}}; perform BN on the source half batch and target half batch respectively, namely \emph{\textbf{BN3}}. For the second data feeding method, BN is performed on the whole batch from each domain, namely \emph{\textbf{BN4}}.

The evaluation results of the above four types of BN are shown in Table~\ref{tab:3}. On the one hand, the average results of all \emph{\textbf{BN1}}-\emph{\textbf{3}} are higher than \emph{\textbf{BN4}}. This proves that it is better to combine data from both source and target domains in one batch than batch them separately. Therefore, it is important for the training of DANN to ensure the guiding gradient signal comes from both source and target domains at each training step. On the other hand, we can also find the average results of both \emph{\textbf{BN1}}-\emph{\textbf{2}} are better than \emph{\textbf{BN3}}. The main difference between \emph{\textbf{BN1}}-\emph{\textbf{2}} and \emph{\textbf{BN3}} is whether the input features for $G_l$ is performed BN on the whole batch (\emph{\textbf{BN1}}-\emph{\textbf{2}}) or on the source and target half separately (\emph{\textbf{BN3}}). This result presents that it is more suitable to feed the language classifier with features performed BN on the entire batch. Besides, it is also worth noting that when training on the smaller database of RECOLA (1,308 utterances), the results of all four settings don’t show significant difference. Therefore,  this study empirically suggests that \emph{\textbf{BN1}} or \emph{\textbf{BN2}} is a more recommended way for BN of features, when training the DANN model on a larger corpus.
\begin{table}[th]
  \vspace{-0.2cm} 
  \caption{UAR (\%)  for four different ways of data combination for batch normalization.}
  \label{tab:3}
  \centering
  \renewcommand\tabcolsep{1.3pt}
  \scalebox{0.865}{
  \begin{tabular}{cccccc}
    \toprule
    & \multicolumn{2}{c}{\textbf{Rec2Iem}} & \multicolumn{2}{c}{\textbf{Iem2Rec}} & \\
    \cmidrule(lr){2-3} \cmidrule(lr){4-5}
    \textbf{model} & \textbf{arousal} &\textbf{valence} & \textbf{arousal} & \textbf{valence} & \textbf{average} \\
    \midrule
    \textbf{BN1} & $71.99(0.33)$ & $54.54(0.77)$ & $\bm{63.18}(\bm{0.32})$ &  $\bm{61.43}(\bm{1.38})$ & $\bm{62.78}$ \\
    \textbf{BN2} & $72.22(0.18)$ & $54.99(0.84)$ & $\bm{62.34}(\bm{0.92})$ &  $\bm{61.37}(\bm{1.15})$ & $\bm{62.73}$ \\
    \textbf{BN3} & $72.27(0.39)$ & $54.07(0.42)$ & $61.83(1.03)$ & $58.48(0.78)$ & $61.66$ \\
    \textbf{BN4} & $72.01(0.33)$ & $53.68(1.26)$ & $60.95(0.90)$ & $56.77(2.84)$ & $60.85$ \\
    \bottomrule
  \end{tabular}}
  
  \vspace{-0.2cm} 
\end{table}

\section{Conclusions}
In this paper, we propose a DANN-based approach for cross-lingual SER\@. Our method works in a completely unsupervised  way, where unlabeled target language data is required only. Experimental results show that our method enables the model to focus on the emotion related information, while ignoring the variations between different languages. Moreover, we explore the impact of batch normalization on training DANN models and suggest two practically optimal ways of data combination  for batch normalization. For further work, we plan to add more corpora from other languages for the cross-lingual SER task.


\section{Acknowledgements}
This work is supported by joint research fund of National Natural Science Foundation of China - Research Grant Council of Hong Kong (NSFC-RGC) (61531166002, N\_CUHK404/15).

\bibliographystyle{IEEEtran}

\bibliography{refbib}

\end{document}